\documentclass[11pt,a4paper,dvips]{article}

\usepackage{colortbl}
\usepackage{longtable}
\usepackage{pstricks, pst-node, pst-tree}
\usepackage{graphicx}
\usepackage{amssymb}
\usepackage[fleqn]{amsmath}
\usepackage{amsthm}

\numberwithin{equation}{section}

\definecolor{avandgreen100}{RGB}{140,198,63}
\definecolor{avandgreen060}{RGB}{186,221,140}
\definecolor{avandgreen020}{RGB}{232,244,217}
\definecolor{avandblue100}{RGB}{0,68,124}
\definecolor{avandblue060}{RGB}{102,143,176}
\definecolor{avandblue020}{RGB}{204,218,229}
\definecolor{avandblack100}{RGB}{0,0,0}
\definecolor{avandblack060}{RGB}{102,102,102}
\definecolor{avandblack020}{RGB}{102,102,102}
\definecolor{avandwhite}{RGB}{255,255,255}

\begin{document}

\title{Using rational numbers to key nested sets}
\author{Dan Hazel\thanks{Technology One.  Email: \texttt{dan\_hazel@technologyonecorp.com}}}
\date{}
\maketitle

%\newpage

\begin{abstract}
This report details the generation and use of tree node ordering keys
in a single relational database table.  The keys for each node are calculated
from the keys of its parent, in such a way that the sort order places
every node in the tree before all of its descendants and after all
siblings having a lower index.  The calculation from parent keys to
child keys is simple, and reversible in the sense that the keys of every
ancestor of a node can be calculated from that node's keys 
without having to consult the database.

Proofs of the above properties of the key encoding process and of its
correspondence to a finite continued fraction form are provided.
\end{abstract}

%\newpage
\section{Introduction: Nested Sets}

The database of interest uses an encoding of nested sets or nested intervals 
to maintain the hierarchical structure of its data.  

\subsection{Nested Sets: $LV$s and $RV$s}

Earlier revisions of the database used the left values and right values 
described by Celko \cite{983361} to key tree nodes.  

\newcommand{\Child}[1]{{\psset{linecolor=avandblue060}\TR{#1}}}
\newcommand{\bb}[1]{{\color{avandblue100}#1}}
\psset{nodesep=3pt, angleA=-90, angleB=90}      
\newsavebox{\graphBox}
\begin{lrbox}{\graphBox}
  \pstree[treemode=D]{\Child{{${}_\bb{1}[\circ \; 1]{}_\bb{2}$}}}{}%
  \pstree[treemode=D]{\Child{{${}_\bb{3}[\circ \; 2]{}_\bb{20}$}}}{%
    \pstree[treemode=D]{\Child{${}_\bb{4}[\circ \; 2 \circ 1]{}_\bb{5}$}}{}
    \pstree[treemode=D]{\Child{${}_\bb{6}[\circ \; 2 \circ 2]{}_\bb{7}$}}{}
    \pstree[treemode=D]{\Child{${}_\bb{8}[\circ \; 2 \circ 3]{}_\bb{9}$}}{}
    \pstree[treemode=D]{\Child{${}_\bb{10}[\circ \; 2 \circ 4]{}_\bb{17}$}}{%
      \Child{${}_\bb{11}[\circ \; 2 \circ 4 \circ 1]{}_\bb{12}$}
      \Child{${}_\bb{13}[\circ \; 2 \circ 4 \circ 2]{}_\bb{14}$}
      \Child{${}_\bb{15}[\circ \; 2 \circ 4 \circ 3]{}_\bb{16}$}}
    \pstree[treemode=D]{\Child{${}_\bb{18}[\circ \; 2 \circ 5]{}_\bb{19}$}}{}}
  \pstree[treemode=D]{\Child{{${}_\bb{21}[\circ \; 3]{}_\bb{22}$}}}{}
\end{lrbox}

\begin{figure}[h]
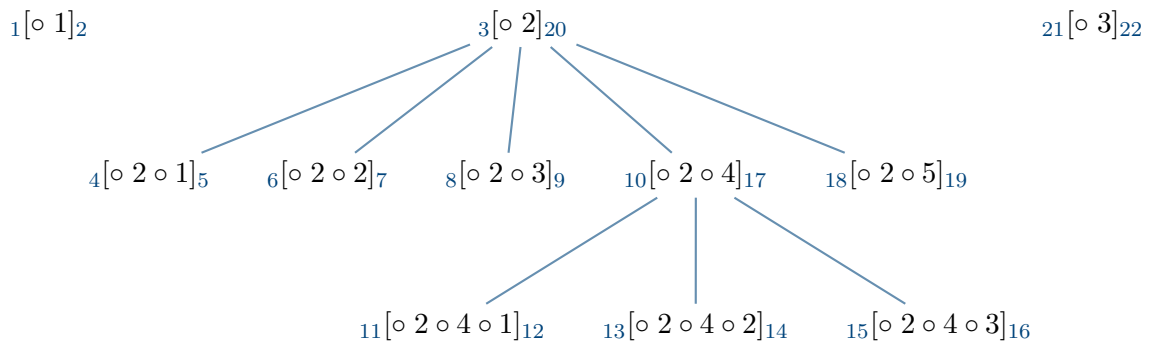

\begin{center}
\leavevmode
\usebox{\graphBox}
\end{center}
\caption{{Node Keys: $LV$ and $RV$.  Each node is shown with $LV$ and $RV$: ${}_\bb{LV}[Tree Position]{}_\bb{RV}$.}}
\label{lvrvtree}
\end{figure}

The nodes of trees keyed in this way are amenable to hierarchy 
painting predicates that are simple enough to be expressed in SQL.

For each of my ancestor nodes,
\begin{equation}
\label{eqnlvrvancpred}
LV_{anc} < LV_{me} < RV_{me} < RV_{anc}
\end{equation}

And so of course for each of my descendant nodes,
\begin{equation}
\label{eqnlvrvdescpred}
LV_{me} < LV_{desc} < RV_{desc} < RV_{me}
\end{equation}

These predicates are useful in the database for determining the
ancestor nodes of a given node, for determining the descendant nodes
of a given node, and most importantly, for imposing an order of
display of a result set that relates directly to the tree.

The immediate problem with this approach is that node insertion 
eventually requires subtrees to the right to be re-encoded.

Consider for example in the tree shown in Figure \ref{lvrvtree},
inserting another node $[\circ \; 2 \circ 2 \circ 1]$, under $[\circ
\; 2 \circ 2]$.  To make room for the keys of the new node that must
satisfy the above predicates, all $LV$s and $RV$s in the tree having
values greater than or equal to $7$ must be incremented.

\subsection{Rational numbers as nodes keys}
\label{usingrationalstokeynodesinstead}

Using rationals as keys obviates the problem with insertion into
nested sets keyed on integer $LV$s and $RV$s.  Within data
representation limits, there will always be an arbitrary number of
rational values between the rational key of any given node and the
rational key of its next closest sibling.  These rational values are
available as keys for the descendants of that node.

\newcommand{\bluefrac}[2][]{\color{avandblue100}\frac{#1}{#2}}
\psset{nodesep=3pt, angleA=-90, angleB=90}      
\newsavebox{\graphCFBox}
\begin{lrbox}{\graphCFBox}
  \pstree[treemode=D]{\Child{{$[\circ \; 1]\bluefrac[1]{1}$}}}{}%
  \pstree[treemode=D]{\Child{{$[\circ \; 2]\bluefrac[2]{1}$}}}{%
    \pstree[treemode=D]{\Child{$[\circ \; 2 \circ 1]\bluefrac[5]{2}$}}{}
    \pstree[treemode=D]{\Child{$[\circ \; 2 \circ 2]\bluefrac[8]{3}$}}{}
    \pstree[treemode=D]{\Child{$[\circ \; 2 \circ 3]\bluefrac[11]{4}$}}{}
    \pstree[treemode=D]{\Child{$[\circ \; 2 \circ 4]\bluefrac[14]{5}$}}{%
      \Child{$[\circ \; 2 \circ 4 \circ 1]\bluefrac[31]{11}$}
      \Child{$[\circ \; 2 \circ 4 \circ 2]\bluefrac[48]{17}$}
      \Child{$[\circ \; 2 \circ 4 \circ 3]\bluefrac[65]{23}$}}
    \pstree[treemode=D]{\Child{$[\circ \; 2 \circ 5]\bluefrac[17]{6}$}}{}}
  \pstree[treemode=D]{\Child{{$[\circ \; 3]\bluefrac[3]{1}$}}}{}
\end{lrbox}

\begin{figure}[h]
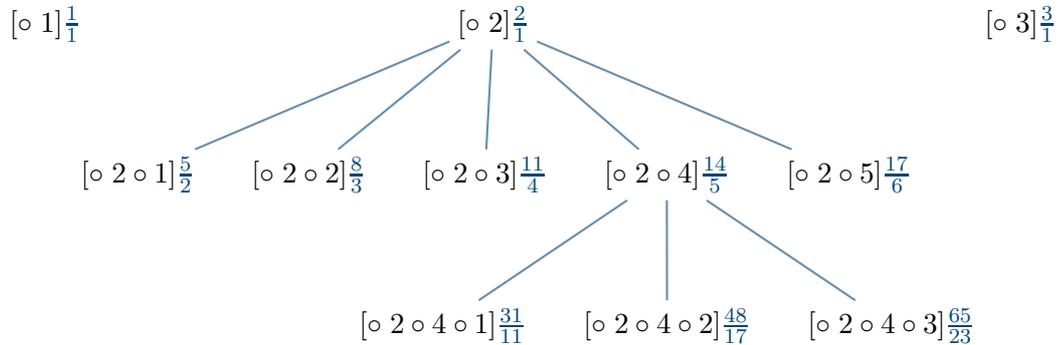

\begin{center}
\leavevmode
\usebox{\graphCFBox}
\end{center}
\caption{Node Keys: Numerators ($nv$) and Denominators ($dv$). Each node is shown as: $[Tree Position]\bluefrac[nv]{dv}$.}
\label{nvdvtree}
\end{figure}

Notice in Figure \ref{nvdvtree} for example, the rational key for each
node under $[\circ \; 2]$, falls strictly between $\frac{2}{1}$ and
$\frac{3}{1}$.  Similarly, the rational key for each node under
$[\circ \; 2 \circ 4]$, falls strictly between $\frac{14}{5}$ and
$\frac{17}{6}$.

\subsection{Determining nv and dv}
A finite continued fraction encoding of tree position provides a unique
rational for each position in the tree.  For example, the 3rd child of
the 4th child of the 2nd top level position, is encoded as
\begin{equation}
\label{eqncfexample}
\mathbb{Q}_{[\circ 2 \circ 4 \circ 3]} 
  = 2+\cfrac{1}{1+\cfrac{1}{4+\cfrac{1}{1+\cfrac{1}{3}}}}
  = \frac{65}{23}
\end{equation}

This is a simple or regular finite continued fraction with an odd
number of terms, where every even numbered term is unity.  A
definition of this $\mathbb{Q}_{[seq]}$ notation is provided in
Section \ref{proofs} on page \pageref{proofs}.

This encoding is very similar to the encoding presented by Tropashko
\cite{DBLP:journals/corr/cs-DB-0402051}.  The Tropashko encoding of
the above tree position is
\begin{equation}
\text{Trop}_{[\circ 2 \circ 4 \circ 3]} 
  =2+\cfrac{1}{4+\cfrac{1}{3}}
  =\frac{29}{13}
\end{equation}

\begin{minipage}{\textwidth}
One problem with the Tropashko encoding, when taken as a finite
continued fraction, is that the rational result for any first child is
the same as that for any next sibling.  For example
\begin{equation}
\text{Trop}_{[\circ 2 \circ 4 \circ 3 \circ 1]} 
  =2+\cfrac{1}{4+\cfrac{1}{3+\cfrac{1}{1}}}
  =2+\cfrac{1}{4+\cfrac{1}{3+1}}
  =2+\cfrac{1}{4+\cfrac{1}{4}}
  =\text{Trop}_{[\circ 2 \circ 4 \circ 4]} 
\end{equation}
\end{minipage}

Tropashko recognises this and ensures that the encodings are always
nicely expressed as their associated continued fraction descendant
functions.  The constraint placed on the form of our encoding allows
us to use simple finite continued fractions instead.

The more difficult problem with the Tropashko encoding is that although 
the rational key of each descendant lies strictly between the keys of its parent
and its parent's next sibling, they do not order monotonically.
Every second row reverses the ordering.

\begin{minipage}{\textwidth}
For example, on level 3 of the Tropashko tree
\begin{equation}
\label{level3}
  2+\cfrac{1}{4+\cfrac{1}{3}} < 
  2+\cfrac{1}{4+\cfrac{1}{4}} < 
  2+\cfrac{1}{4+\cfrac{1}{5}}
\end{equation}
that is,
\begin{equation*}
  \frac{29}{18} < 
  \frac{38}{17} < 
  \frac{47}{21} 
\end{equation*}

However, on level 4 of the tree
\begin{equation}
\label{level4}
  2+\cfrac{1}{4+\cfrac{1}{3+\cfrac{1}{3}}} > 
  2+\cfrac{1}{4+\cfrac{1}{3+\cfrac{1}{4}}} > 
  2+\cfrac{1}{4+\cfrac{1}{3+\cfrac{1}{5}}}
\end{equation}
that is,
\begin{equation*}
  \frac{96}{43} >
  \frac{125}{56} > 
  \frac{154}{69} 
\end{equation*}

Notice the difference in direction between the inequalities in
\eqref{level4} and those in \eqref{level3}.  This makes the order by
clause over a database result set extremely difficult to phrase.  It
could be said that our encoding has reestablished monotonicity by
leaving out every second row.
\end{minipage}

Tropashko recognises the monotonicity problem in his later paper
\cite{DBLP:journals/sigmod/Tropashko05}.  A workaround using reverse
continued fractions, different from our encoding, is suggested there.
However imposing an order where a child key is greater than a parent key,
corresponds to keys for earlier siblings being greater than more
recent siblings.

\section{Sibling quadruples}

\subsection{Next sibling numerator and denominator: $snv$ and $sdv$}

We find it expedient to store with each node, not only the $nv$ and
$dv$ that define its rational key, but also the numerator and
denominator of the node's next sibling, $snv$ and $sdv$.  These values
will be used when searching for a nodes descendants as is hinted in
Section \ref{usingrationalstokeynodesinstead}.  They are also seminally useful
when determining the $nv$ and $dv$ of an inserted child node and, as explained in Section \ref{transformations}, when relocating subtrees.

The choice to keep $snv$ and $sdv$ with $nv$ and $dv$ diverges from common practice in the continued fractions literature of keeping the parent keys.
It is our constraint on the form of continued fractions we employ that makes $snv$ and $sdv$ the preferred associated pair.  The matrices we associate with tree 
nodes should not be confused with those often used in reasoning about continued fractions.

The next sibling of $[\circ \; 2 \circ 4 \circ 3]$ is $[\circ \; 2 \circ 4 \circ 4]$.
\begin{align*}
\mathbb{Q}_{[\circ 2 \circ 4 \circ 3]}& =\frac{65}{23}\\
\mathbb{Q}_{[\circ 2 \circ 4 \circ 4]}& =2+\cfrac{1}{1+\cfrac{1}{4+\cfrac{1}{1+\cfrac{1}{4}}}} = \frac{82}{29}
\end{align*}

\begin{figure}[!h]
\begin{center}
\leavevmode
\begin{align*}
\quad& \text{Tree position} & \mathbf{nv} && \mathbf{dv} && \mathbf{snv} && \mathbf{sdv}&\quad\\
\quad& [\circ \; 2] & 2 && 1 && {\color{red}\mathbf{3}} && {\color{red}\mathbf{1}}&\quad\\
\quad& [\circ \; 2 \circ 1] & 5 && 2 && 8 && 3&\quad\\
\quad& [\circ \; 2 \circ 2] & 8 && 3 && 11 && 4&\quad\\
\quad& [\circ \; 2 \circ 3] & 11 && 4 && 14 && 5&\quad\\
\quad& [\circ \; 2 \circ 4] & 14 && 5 && {\color{red}\mathbf{17}} && {\color{red}\mathbf{6}}&\quad\\
\quad& [\circ \; 2 \circ 4 \circ 1] & 31 && 11 && 48 && 17&\quad\\
\quad& [\circ \; 2 \circ 4 \circ 2] & 48 && 17 && 65 && 23&\quad\\
\quad& [\circ \; 2 \circ 4 \circ 3] & 65 && 23 && 82 && 29&\quad
\end{align*}
\end{center}
\caption{Some example keys}
\label{whysnvsdv}
\end{figure}

Figure \ref{whysnvsdv} shows the $nv$, $dv$, $snv$ and $sdv$ of some
of the nodes in our example tree.  Notice that when determining the
$nv$ and $dv$ of the node $[\circ \; 2 \circ 4 \circ 3]$, we could
either perform the continued fraction calculation shown at
\eqref{eqncfexample}, or we could use the values on the parent row,
$[\circ \; 2 \circ 4]$.  Adding $nv$ to $snv$ gives the numerator of
the first child.  Adding $dv$ to $sdv$ gives the denominator of the
first child.  Adding $nv$ to $3 \times snv$ gives the numerator of the
third child.  Adding $dv$ to $3 \times sdv$ gives the denominator of
the third child.

In general, given the $nv$, $dv$, $snv$ and $sdv$ of a parent node $p$, we can determine the $nv$ and $dv$ of its $c^{th}$ child as follows:
\begin{align}
\label{eqnnvdvsumbegin}
nv_{c} &= nv_{p} + c \times snv_{p}\\
\label{eqnnvdvsumbegindenom}
dv_{c} &= dv_{p} + c \times sdv_{p}
\end{align}

A proof of \eqref{eqnnvdvsumbegin} and \eqref{eqnnvdvsumbegindenom} is provided in Section \ref{proofs}.

Since the next sibling of the $c^{th}$ child of node $p$, is the ${(c + 1)}^{th}$ child of node $p$, it follows that
\begin{align}
snv_{c} &= nv_{p} + (c + 1) \times snv_{p}\\
\label{eqnnvdvsumend}
sdv_{c} &= dv_{p} + (c + 1) \times sdv_{p}
\end{align}

A concrete example from values in Figure \ref{whysnvsdv} is
\begin{align*}
65 &= 14 + 3 \times 17\\
23 &= 5 + 3 \times 6 \\
82 &= 14 + (3 + 1) \times 17\\
29 &= 5 + (3 + 1) \times 6
\end{align*}

That is:
\begin{align*}
nv_{[\circ 2 \circ 4 \circ 3]} &= nv_{[\circ 2 \circ 4]} + 3 \times snv_{[\circ 2 \circ 4]} \\
dv_{[\circ 2 \circ 4 \circ 3]} &= dv_{[\circ 2 \circ 4]} + 3 \times sdv_{[\circ 2 \circ 4]} \\
snv_{[\circ 2 \circ 4 \circ 3]} &= nv_{[\circ 2 \circ 4]} + (3 + 1) \times snv_{[\circ 2 \circ 4]} \\
sdv_{[\circ 2 \circ 4 \circ 3]} &= dv_{[\circ 2 \circ 4]} + (3 + 1) \times sdv_{[\circ 2 \circ 4]}
\end{align*}

\subsection{Tree hierarchy predicates}

The predicates that can be used to filter ancestors of a given node or
descendants of a given node are not quite as simple as those available when
using $LV$s and $RV$s to key nodes.  See Predicates
\eqref{eqnlvrvancpred} and \eqref{eqnlvrvdescpred}.

For the encoding presented here, if a node, $me$, has keys, 
$(nv_{me}, dv_{me}, snv_{me}, sdv_{me})$,
then a node, $anc$, with keys,
$(nv_{anc}, dv_{anc}, snv_{anc}, sdv_{anc})$,
is an ancestor of $me$ \mbox{if{f}}:
\begin{equation}
\label{eqnancpred}
\frac{nv_{anc}}{dv_{anc}} 
< \frac{nv_{me}}{dv_{me}} 
< \frac{snv_{anc}}{sdv_{anc}}
\end{equation}

and a node, $desc$, with keys,
$(nv_{desc}, dv_{desc}, snv_{desc}, sdv_{desc})$,
is a descendant of $me$ \mbox{if{f}}:
\begin{equation}
\label{eqndescpred}
\frac{nv_{me}}{dv_{me}} 
< \frac{nv_{desc}}{dv_{desc}} 
< \frac{snv_{me}}{sdv_{me}}
\end{equation}

In practice, the predicate to filter ancestors is not used.  
This is because with the continued fractions encoding, the keys of all 
ancestors of a given node, $\eta$, can be calculated from the 
$nv$ and $dv$ keys of $\eta$.
There is rarely a need to use the inequalities of \eqref{eqnancpred} 
to test whether a node is 
an ancestor of another node.

The source code for a SQL Server 2005 function to return the ancestors 
of a node indicated by argument numerator and denominator is provided in 
Figure \ref{figgetancestorsastable} on page \pageref{figgetancestorsastable}.
This algorithm performs a simple root to leaf walk through the continued 
fraction encoding.

On the other hand, while in principle, calculation of descendants is
possible, the \eqref{eqndescpred} inequalities are used to filter
descendant subtree searches since we would have to go to the database
anyway to ask how many children each descendant has.

\newsavebox{\codebox}
\begin{lrbox}{\codebox}
\begin{minipage}{\textwidth}
\begin{verbatim}
create function getAncestorsAsTable(
    @numerator bigint,
    @denominator bigint
) returns @ancestortable table (nv bigint not null, dv bigint not null)
as
begin
    declare @ancnv bigint
    declare @ancdv bigint
    declare @ancsnv bigint
    declare @ancsdv bigint
    
    declare @div bigint
    declare @mod bigint
    
    set @ancnv = 0
    set @ancdv = 1
    set @ancsnv = 1
    set @ancsdv = 0
    
    while @numerator > 0 and @denominator > 0
    begin
        set @div = @numerator / @denominator
        set @mod = @numerator % @denominator
        
        set @ancnv = @ancnv + @div * @ancsnv
        set @ancdv = @ancdv + @div * @ancsdv
        set @ancsnv = @ancnv + @ancsnv
        set @ancsdv = @ancdv + @ancsdv
        
        insert into @ancestortable (nv, dv) values (@ancnv, @ancdv)
        
        set @numerator = @mod
        if @numerator <> 0
        begin
            set @denominator = @denominator % @mod
            if @denominator = 0
            begin
                set @denominator = 1
            end
        end
    end
    return
end
\end{verbatim}
\end{minipage}
\end{lrbox}

\begin{figure}
\usebox{\codebox}
\caption{SQL Server 2005 function to return a table of 
ancestor keys when passed a numerator and denominator}
\label{figgetancestorsastable}
\end{figure}

\section{Transformations}
\label{transformations}
	
\subsection{Offspring transformation}

\begin{minipage}{\textwidth}
If we draw the quadruple: $(nv, dv, snv, sdv)$,
as a $2 \times 2$ matrix:
\begin{equation}
\begin{bmatrix}
nv&snv\\
dv&sdv
\end{bmatrix}
\end{equation}
Then 
\begin{align}
\mathbb{M}_{[\circ 2 \circ 4]} & = \Bigl [
\begin{smallmatrix}
14&17\\
5&6
\end{smallmatrix}
\Bigr ] \\[-0.1cm]
\intertext{And}
\mathbb{M}_{[\circ 2 \circ 4 \circ 3]} & = \Bigl [
\begin{smallmatrix}
65&82\\
23&29
\end{smallmatrix}
\Bigr ] \\[-0.5cm]
\intertext{And}
\label{eqnmatrixexample}
\mathbb{M}_{[\circ 2 \circ 4 \circ 3]} & = \mathbb{M}_{[\circ 2 \circ 4]} \Bigl [
\begin{smallmatrix}
1&1\\
3&(3+1)
\end{smallmatrix}
\Bigr ]
\end{align}

The equality shown in \eqref{eqnmatrixexample} is just an application
of equations \eqref{eqnnvdvsumbegin} through \eqref{eqnnvdvsumend}.
\end{minipage}

\begin{minipage}{\textwidth}
  Because of equations \eqref{eqnnvdvsumbegin} through
  \eqref{eqnnvdvsumend}, the matrix corresponding to each node in our
  tree is built of a product of transformations that lead back to the
  root of the tree.  For example:
\begin{equation}
\mathbb{M}_{[\circ 2 \circ 4 \circ 3]} = \Bigl [
\begin{smallmatrix}
65&82\\
23&29
\end{smallmatrix}
\Bigr ]
=
\Bigl [
\begin{smallmatrix}
0&1\\
1&0
\end{smallmatrix}
\Bigr ]
\Bigl [
\begin{smallmatrix}
1&1\\
2&(2+1)
\end{smallmatrix}
\Bigr ]
\Bigl [
\begin{smallmatrix}
1&1\\
4&(4+1)
\end{smallmatrix}
\Bigr ]
\Bigl [
\begin{smallmatrix}
1&1\\
3&(3+1)
\end{smallmatrix}
\Bigr ]
\end{equation}
\end{minipage}

\begin{minipage}{\textwidth}
An important observation in regard to performing calculations within 
the database is that the determinant of each of the factor matrices is 
either $-1$ or $1$.
\begin{align}
\Bigl |
\begin{smallmatrix}
0&1\\
1&0
\end{smallmatrix}
\Bigr |
& = 0-1 = -1\\
\Bigl |
\begin{smallmatrix}
1&1\\
c&(c+1)
\end{smallmatrix}
\Bigr |
& = 1 \times (c+1) - 1 \times c = 1\\[-0.1cm]
\end{align}

And so, the determinant of the product is
\begin{equation}
\text{det} \bigl ( \mathbb{M}_{[\circ 2 \circ 4 \circ 3]} \bigr ) = -1 \times 1 \times 1 \times 1 = -1
\end{equation}
\end{minipage}

This property is used in Section \ref{proofs} to show that each pair of $nv$ and $dv$ are relatively prime.
This is as good a normal form as any.

\begin{minipage}{\textwidth}
Also, since the determinant of the matrix of each node in our tree is $-1$,
the inverse of any matrix,
$ \Bigl [ \begin{smallmatrix}nv&snv\\dv&sdv\end{smallmatrix} \Bigr ] $
is given by
\begin{align}
\Bigl [
\begin{matrix}
nv&snv\\
dv&sdv
\end{matrix}
\Bigr ] ^ {-1}
&=
\frac{1}{\bigl |
\begin{smallmatrix}
nv&snv\\
dv&sdv
\end{smallmatrix}
\bigr | } \times
\Bigl [
\begin{matrix}
sdv&-snv\\
-dv&nv
\end{matrix}
\Bigr ] \\
\quad & = 
\Bigl [
\begin{matrix}
-sdv&snv\\
dv&-nv
\end{matrix}
\Bigr ]
\end{align}

Which means that inverse transformations can be calculated in the
database without the need to leave integer arithmetic.
Inverse transformations are important to the process of moving subtrees.
\end{minipage}

\subsection{Moving subtrees}

If it is required to move a subtree from under the $n$th child of the
node with matrix $p_0$ to under the $m$th child of the node with
matrix $p_1$, this can be achieved using the relatively immediate
availability of the inverses of the matrices.  Say an arbitrary node
in that subtree is given by matrix $M_0$, then there must be a $\varphi$ such
that
\begin{align}
p_0 \times \Bigl [
\begin{smallmatrix}
1&1\\
n&n+1
\end{smallmatrix}
\Bigr ]
\times \varphi
& =
M_0\\
\Bigl [
\begin{smallmatrix}
1&1\\
n&n+1
\end{smallmatrix}
\Bigr ]
\times \varphi
& =
p_0 ^ {-1} \times M_0\\
\varphi
& =
\Bigl [
\begin{smallmatrix}
1&1\\
n&n+1
\end{smallmatrix}
\Bigr ] ^ {-1}
\times p_0 ^ {-1} \times M_0\\
\Bigl [
\begin{smallmatrix}
1&1\\
m&m+1
\end{smallmatrix}
\Bigr ]
\times \varphi
& =
\Bigl [
\begin{smallmatrix}
1&1\\
m&m+1
\end{smallmatrix}
\Bigr ]
\times \Bigl [
\begin{smallmatrix}
1&1\\
n&n+1
\end{smallmatrix}
\Bigr ] ^ {-1}
\times p_0 ^ {-1} \times M_0\\
\label{eqnexfinalrelocation}
p_1 \times \Bigl [
\begin{smallmatrix}
1&1\\
m&m+1
\end{smallmatrix}
\Bigr ]
\times \varphi
& =
p_1 \times \Bigl [
\begin{smallmatrix}
1&1\\
m&m+1
\end{smallmatrix}
\Bigr ]
\times \Bigl [
\begin{smallmatrix}
1&1\\
n&n+1
\end{smallmatrix}
\Bigr ] ^ {-1}
\times p_0 ^ {-1} \times M_0
\end{align}

The left hand side of the equality \eqref{eqnexfinalrelocation}
expresses the relocation of the subtree to the $m^{th}$ child of
$p_1$.

Simplifying:
\begin{equation}
\Bigl [
\begin{smallmatrix}
1&1\\
m&m+1
\end{smallmatrix}
\Bigr ]
\times \Bigl [
\begin{smallmatrix}
1&1\\
n&n+1
\end{smallmatrix}
\Bigr ] ^ {-1} =
\Bigl [
\begin{smallmatrix}
1&0\\
(m - n)&1
\end{smallmatrix}
\Bigr ]
\end{equation}

Restating, when a subtree identified as the descendants of the $n$th
child of the node with matrix $p_0$ is relocated to the subtree
identified as the descendants of the $m$th child of the node with
matrix $p_1$, any descendant node having matrix $M_0$ before the
relocation, will have matrix $M_1$ after, where $M_1$ is given by
\begin{equation}
M_1 =
p_1 \times \Bigl [
\begin{smallmatrix}
1&0\\
(m - n)&1
\end{smallmatrix}
\Bigr ]
\times p_0 ^ {-1} \times M_0
\end{equation}

\section{Properties of the encoding}
\label{proofs}

\newtheorem{theorem}{Theorem}[section]
\newtheorem{definition}[theorem]{Definition}

\begin{definition}
\label{defnQ}

Our encoding uses a simple or regular finite continued fraction with
an odd number of terms, where every even numbered term is unity.
It allows the terms to range over positive real numbers, $\mathbb{R}^+$, to make the proofs easier,
though in general use, the terms are strictly in $\mathbb{N}$.
\begin{equation}
\begin{split}
\forall & N_1, N_2, \dotsb, N_n \in \mathbb{R}^+ \\
& \mathbb{Q}_{[\circ N_1 \circ N_2 \circ \dotsb \circ N_n]} 
  \overset{\text{\tiny def}}{=} N_1 + \cfrac{1}{1+\cfrac{1}{N_2+\cfrac{1}{1+ \cfrac{1}{\ddots + \cfrac{\vdots}{\cfrac{1}{1 + \cfrac{1}{N_n}}}}}}}
\end{split}
\end{equation}

It is accepted and used without proof that:
\begin{equation}
\begin{split}
\forall & N_1, \dotsb, N_n, \delta, \varepsilon \in \mathbb{R}^+ \\
& \mathbb{Q}_{[\circ N_1 \circ \dotsb \circ N_n \circ \delta \circ \varepsilon]} = 
\mathbb{Q}_{\left [\circ N_1 \circ \dotsb \circ N_n \circ \left (\delta + \cfrac{1}{1 + \cfrac{1}{\varepsilon}} \right ) \right]}
\end{split}
\end{equation}

\end{definition}

\begin{minipage}{\textwidth}

Below, $\prod_{k=1}^m M_k$ denotes the product of a sequence of $2 \times 2$ matrices in the order $M_1 M_2 \dotsb M_m$.

\begin{theorem}
\label{gcdQ}
The generated keys of tree nodes are in their lowest terms.
That is, each numerator and denominator pair has no common divisors.\\[2mm]

FOR ALL
\begin{align*}
& m \in \mathbb{N}\\
\intertext{AND}
& nv_m, dv_m, snv_m, sdv_m, N_1, ..., N_m \in \mathbb{N}
\end{align*}

PROVIDED
\begin{equation}
\label{seqdefn}
\Bigl [
\begin{smallmatrix} nv_m & snv_m\\ dv_m & sdv_m \end{smallmatrix}
\Bigr ]
=
\Bigl [
\begin{smallmatrix} 0 & 1\\ 1 & 0 \end{smallmatrix}
\Bigr ]
\prod_{k=1}^{m} 
\Bigl [
\begin{smallmatrix} 1 & 1\\ N_{k} & (N_{k} + 1) \end{smallmatrix}
\Bigr ]
\end{equation}

HOLDS
\begin{align}
\label{gcdone}
& \gcd(nv_m,dv_m) = 1\\
\label{gcdtwo}
& \gcd(snv_m,sdv_m) = 1\\
\label{gcdthree}
& \gcd(nv_m,snv_m) = 1\\
\label{gcdfour}
& \gcd(dv_m,sdv_m) = 1
\end{align}

\end{theorem}

\end{minipage}

\subsection*{Proof of Theorem \ref{gcdQ}}

Consider first, for $nv_m, dv_m, snv_m, sdv_m, N_1, ..., N_m \in \mathbb{N}$:
\begin{equation}
\label{detequalsunity}
\begin{split}
nv_m \times sdv_m - dv_m \times snv_m
& = 
\Bigl |
\begin{smallmatrix} nv_m & snv_m\\ dv_m & sdv_m \end{smallmatrix}
\Bigr | & \text{by definition of determinant}\\
& =
\text{det} \Bigl (
\Bigl [
\begin{smallmatrix} 0 & 1\\ 1 & 0 \end{smallmatrix}
\Bigr ]
\prod_{k=1}^{m} 
\Bigl [
\begin{smallmatrix} 1 & 1\\ N_{k} & (N_{k} + 1) \end{smallmatrix}
\Bigr ]
\Bigr ) & \text{using hypothesis \eqref{seqdefn}}\\
& =
\Bigl |
\begin{smallmatrix} 0 & 1\\ 1 & 0 \end{smallmatrix}
\Bigr | \times
\prod_{k=1}^{m} 
\Bigl |
\begin{smallmatrix} 1 & 1\\ N_{k} & (N_{k} + 1) \end{smallmatrix}
\Bigr | & \text{(scalar $\prod$ now)}\\
& = -1 \times \prod_{k=1}^{m} 1 & \text{by calculation}\\
& = -1 &
\end{split}
\end{equation}

For all $a \in \mathbb{N}$ 
such that $a$ is a divisor of both $nv_m$ and $dv_m$
there must be $b, c \in \mathbb{N}$ such that $nv_m = a \times b$ 
and $dv_m = a \times c$.

In which case,
\begin{equation}
\begin{split}
-1 & = nv_m \times sdv_m - dv_m \times snv_m & \text{using \eqref{detequalsunity}} \\
& = a \times b \times sdv_m - a \times c \times snv_m & \\
& = a \times (b \times sdv_m - c \times snv_m) &
\end{split}
\end{equation}

It follows that for all $a \in \mathbb{N}$ 
such that $a$ is a divisor of both $nv_m$ and $dv_m$,
$a = 1$, as required for \eqref{gcdone}.

The other $\gcd$ results \eqref{gcdtwo}, \eqref{gcdthree} and \eqref{gcdfour}
are proven similarly, also using \eqref{detequalsunity}.

$\qed$ Theorem \ref{gcdQ}.\\[2mm]

\begin{minipage}{\textwidth}

\begin{theorem}
\label{sequencetoQ}
The generated key pair $\cfrac{nv}{dv}$ of each tree node, is unique to the tree.  
Uniqueness rests on well known uniqueness properties of (carefully constrained) continued fractions.\\[2mm]

FOR ALL
\begin{align*}
& m \in \mathbb{N}\\
\intertext{AND}
& nv_m, dv_m, snv_m, sdv_m, N_1, ..., N_m \in \mathbb{N}
\end{align*}

PROVIDED
\begin{equation}
\label{seqdefntwo}
\Bigl [
\begin{smallmatrix} nv_m & snv_m\\ dv_m & sdv_m \end{smallmatrix}
\Bigr ]
=
\Bigl [
\begin{smallmatrix} 0 & 1\\ 1 & 0 \end{smallmatrix}
\Bigr ]
\prod_{k=1}^{m} 
\Bigl [
\begin{smallmatrix} 1 & 1\\ N_{k} & (N_{k} + 1) \end{smallmatrix}
\Bigr ]
\end{equation}

HOLDS
\begin{align}
\label{reqnvdv}
& \frac{nv_m}{dv_m} = \mathbb{Q}_{[\circ N_1 \circ \dotsb \circ N_m]}\\
\label{reqsnvsdv}
& \frac{snv_m}{sdv_m}  = \mathbb{Q}_{[\circ N_1 \circ \dotsb \circ N_{m-1} \circ (N_{m} + 1)]}\\
\label{reqnvdvdelta}
& \forall \delta \in \mathbb{R}^+ \bullet \frac{nv_m + \delta \times snv_m}{dv_m + \delta \times sdv_m} = \mathbb{Q}_{[\circ N_1 \circ \dotsb \circ N_m \circ \delta]}
\end{align}

\end{theorem}
\end{minipage}

\subsection*{Proof of Theorem \ref{sequencetoQ}}

This proof is by induction over $m$, the depth of the tree.
\subsubsection*{Basis: $m = 1$}
\begin{minipage}{\textwidth}
Using Definition \ref{defnQ}:
\begin{equation}
\label{basenv}
\mathbb{Q}_{[\circ N_1]} = N_1
\end{equation}
\begin{equation}
\label{basesnv}
\mathbb{Q}_{[\circ (N_1 + 1)]} = N_1 + 1
\end{equation}
and for all $\delta \in \mathbb{R}^+$,
\begin{equation}
\label{basedelta}
\begin{split}
\mathbb{Q}_{[\circ N_1 \circ \delta]} & = N_1 + \cfrac{1}{1 + \cfrac{1}{\delta}}\\
& = \frac{N_{1} + \delta \times N_{1} + \delta}{\delta + 1}
\end{split}
\end{equation}
\end{minipage}

It is also useful to expand hypothesis \eqref{seqdefntwo}, for $m = 1$:
\begin{equation}
\label{basehyp}
\begin{split}
\Bigl [
\begin{smallmatrix} nv_1 & snv_1\\ dv_1 & sdv_1 \end{smallmatrix}
\Bigr ]
& =
\Bigl [
\begin{smallmatrix} 0 & 1\\ 1 & 0 \end{smallmatrix}
\Bigr ]
\Bigl [
\begin{smallmatrix} 1 & 1\\ N_{1} & (N_{1} + 1) \end{smallmatrix}
\Bigr ] \\
& =
\Bigl [
\begin{smallmatrix} N_{1} & (N_{1} + 1)\\ 1 & 1 \end{smallmatrix}
\Bigr ]
\end{split}
\end{equation}

Then,
\begin{equation*}
\begin{split}
\frac{nv_1}{dv_1} & = \frac{N_{1}}{1} & \text{using \eqref{basehyp}} \\
& = \mathbb{Q}_{[\circ N_1]} & \text{using \eqref{basenv}}
\end{split}
\end{equation*}
as required to show \eqref{reqnvdv} for $n = 1$.

And,
\begin{equation*}
\begin{split}
\frac{snv_1}{sdv_1} & = \frac{N_{1} + 1}{1} & \text{using \eqref{basehyp}} \\
& = \mathbb{Q}_{[\circ (N_1 + 1)]} & \text{using \eqref{basesnv}}
\end{split}
\end{equation*}
as required to show \eqref{reqsnvsdv} for $m = 1$.

And,
\begin{equation*}
\begin{split}
\frac{nv_1 + \delta \times snv_1}{dv_1 + \delta \times sdv_1} & = \frac{N_{1} + \delta \times (N_{1} + 1)}{1 + \delta \times 1} & \text{using \eqref{basehyp}} \\
& = \frac{N_{1} + \delta \times N_{1} + \delta}{\delta + 1} & \text{simplifying}\\
& = \mathbb{Q}_{[\circ N_1 \circ \delta]} & \text{using \eqref{basedelta}}
\end{split}
\end{equation*}
as required to show \eqref{reqnvdvdelta} for $m = 1$.

\subsubsection*{Inductive step}

\begin{minipage}{\textwidth}
It is enough to show that for \[m \in \mathbb{N}\] and 
\[nv_m, dv_m, snv_m, sdv_m, nv_{m+1}, dv_{m+1}, snv_{m+1}, sdv_{m+1}, N_1, ..., N_{m+1} \in \mathbb{N}\]

THAT
\begin{equation}
\label{stepreqnvdvmplusone}
\frac{nv_{m+1}}{dv_{m+1}} = \mathbb{Q}_{[\circ N_1 \circ \dotsb \circ N_{m+1}]}\\
\end{equation}

and
\begin{equation}
\label{stepreqsnvsdvmplusone}
\frac{snv_{m+1}}{sdv_{m+1}}  = \mathbb{Q}_{[\circ N_1 \circ \dotsb \circ N_{m} \circ (N_{m+1} + 1)]}
\end{equation}

and
\begin{equation}
\label{stepreqnvdvdeltamplusone}
\forall \gamma \in \mathbb{R}^+ \bullet \frac{nv_{m+1} + \gamma \times snv_{m+1}}{dv_{m+1} + \gamma \times sdv_{m+1}} = \mathbb{Q}_{[\circ N_1 \circ \dotsb \circ N_{m+1} \circ \gamma]}
\end{equation}

PROVIDED
\begin{equation}
\label{stephypm}
\Bigl [
\begin{smallmatrix} nv_m & snv_m\\ dv_m & sdv_m \end{smallmatrix}
\Bigr ]
=
\Bigl [
\begin{smallmatrix} 0 & 1\\ 1 & 0 \end{smallmatrix}
\Bigr ]
\prod_{k=1}^{m} 
\Bigl [
\begin{smallmatrix} 1 & 1\\ N_{k} & (N_{k} + 1) \end{smallmatrix}
\Bigr ]
\end{equation}

and
\begin{equation}
\label{stepreqnvdvm}
\frac{nv_m}{dv_m} = \mathbb{Q}_{[\circ N_1 \circ \dotsb \circ N_m]}\\
\end{equation}

and
\begin{equation}
\label{stepreqsnvsdvm}
\frac{snv_m}{sdv_m}  = \mathbb{Q}_{[\circ N_1 \circ \dotsb \circ N_{m-1} \circ (N_{m} + 1)]}
\end{equation}

and
\begin{equation}
\label{stepreqnvdvdeltam}
\forall \xi \in \mathbb{R}^+ \bullet \frac{nv_m + \xi \times snv_m}{dv_m + \xi \times sdv_m} = \mathbb{Q}_{[\circ N_1 \circ \dotsb \circ N_m \circ \xi]}
\end{equation}

and
\begin{equation}
\label{stephypmplusone}
\Bigl [
\begin{smallmatrix} nv_{m+1} & snv_{m+1}\\ dv_{m+1} & sdv_{m+1} \end{smallmatrix}
\Bigr ]
=
\Bigl [
\begin{smallmatrix} 0 & 1\\ 1 & 0 \end{smallmatrix}
\Bigr ]
\prod_{k=1}^{m+1} 
\Bigl [
\begin{smallmatrix} 1 & 1\\ N_{k} & (N_{k} + 1) \end{smallmatrix}
\Bigr ]
\end{equation}
\end{minipage}

\subsubsection*{Proof of Inductive step}
It is required to prove \eqref{stepreqnvdvmplusone}, \eqref{stepreqsnvsdvmplusone} and \eqref{stepreqnvdvdeltamplusone}
given the hypotheses \eqref{stephypm}, \eqref{stepreqnvdvm}, \eqref{stepreqsnvsdvm}, \eqref{stepreqnvdvdeltam} and \eqref{stephypmplusone}.

It is useful to first calculate, using \eqref{stephypmplusone} and \eqref{stephypm}:
\begin{equation}
\label{mplusoneexpand}
\begin{split}
\Bigl [
\begin{smallmatrix} nv_{m+1} & snv_{m+1}\\ dv_{m+1} & sdv_{m+1} \end{smallmatrix}
\Bigr ]
& = \Bigl [
\begin{smallmatrix} nv_m & snv_m\\ dv_m & sdv_m \end{smallmatrix}
\Bigr ] \Bigl [
\begin{smallmatrix} 1 & 1\\ N_{m+1} & (N_{m+1} + 1) \end{smallmatrix}
\Bigr ] \\
&= \Bigl [
\begin{smallmatrix} (nv_m + N_{m+1} \times snv_m) & (nv_m + (N_{m+1} + 1) \times snv_m)\\  (dv_m + N_{m+1} \times sdv_m) & (dv_m + (N_{m+1} + 1) \times sdv_m) \end{smallmatrix}
\Bigr ]
\end{split}
\end{equation}

\begin{minipage}{\textwidth}
Consider then hypothesis \eqref{stepreqnvdvdeltam}, choosing $N_{n + 1}$ for $\xi$:
\begin{equation*}
\begin{split}
\mathbb{Q}_{[\circ N_1 \circ \dotsb \circ N_m \circ N_{n + 1}]}
& = \frac{nv_m + N_{n + 1} \times snv_m}{dv_m + N_{n + 1} \times sdv_m} &\\
& = \frac{nv_{m+1}}{dv_{m+1}} & \text{using the final equality at \eqref{mplusoneexpand}}
\end{split}
\end{equation*}
as required to show \eqref{stepreqnvdvmplusone}.
\end{minipage}

\begin{minipage}{\textwidth}
Again using the hypothesis \eqref{stepreqnvdvdeltam}, but this time choosing $N_{n + 1} + 1$ for $\xi$:
\begin{equation*}
\begin{split}
\mathbb{Q}_{[\circ N_1 \circ \dotsb \circ N_m \circ (N_{n + 1} + 1)]}
& = \frac{nv_m + (N_{n + 1} + 1) \times snv_m}{dv_m + (N_{n + 1} + 1) \times sdv_m} &\\
& = \frac{snv_{m+1}}{sdv_{m+1}} & \text{using the final equality at \eqref{mplusoneexpand}}
\end{split}
\end{equation*}
as required to show \eqref{stepreqsnvsdvmplusone}.
\end{minipage}

\begin{minipage}{\textwidth}
Finally, again using the hypothesis \eqref{stepreqnvdvdeltam}, but now choosing 
$N_{n + 1} + \cfrac{1}{1 + \cfrac{1}{\delta}}$ for $\xi$, 
where $\delta \in \mathbb{R}^+$,
\begin{equation*}
%\label{stepreqnvdvdeltamexpand}
\begin{split}
& \mathbb{Q}_{[\circ N_1 \circ \dotsb \circ N_m \circ N_{n + 1} \circ \delta]} &\\
& = \mathbb{Q}_{\left [\circ N_1 \circ \dotsb \circ N_m \circ \left (N_{n + 1} + \cfrac{1}{1 + \cfrac{1}{\delta}} \right ) \right ]} & \text{by Definition \ref{defnQ}}\\
& = \frac{nv_m + \left ( N_{n + 1} + \cfrac{1}{1 + \cfrac{1}{\delta}} \right ) \times snv_m}{dv_m + \left ( N_{n + 1} + \cfrac{1}{1 + \cfrac{1}{\delta}} \right ) \times sdv_m} & \text{by \eqref{stepreqnvdvdeltam}}\\
& = \frac{nv_m + \left ( N_{n + 1} + \cfrac{\delta}{1 + \delta} \right ) \times snv_m}{dv_m + \left ( N_{n + 1} + \cfrac{\delta}{1 + \delta} \right ) \times sdv_m} &\\
& = \frac{nv_m + \cfrac{N_{n + 1} \times \delta + N_{n + 1} + \delta}{1 + \delta} \times snv_m}{dv_m + \cfrac{N_{n + 1} \times \delta + N_{n + 1} + \delta}{1 + \delta} \times sdv_m} &\\
& = \frac{nv_m \times (1 + \delta) + (N_{n + 1} \times \delta + N_{n + 1} + \delta) \times snv_m}{dv_m \times (1 + \delta) + (N_{n + 1} \times \delta + N_{n + 1} + \delta) \times sdv_m} &\\
& = \frac{nv_m + N_{n + 1} \times snv_m + \delta \times (nv_m + (N_{n + 1} + 1) \times snv_m)}{dv_m + N_{n + 1} \times sdv_m + \delta \times (dv_m + (N_{n + 1} + 1) \times sdv_m)} &\\
& = \frac{nv_{m+1} + \delta \times snv_{m+1}}{dv_{m+1} + \delta \times sdv_{m+1}} & \text{using the final equality at \eqref{mplusoneexpand}}
\end{split}
\end{equation*}
which, generalizing $\delta$ to $\gamma$, is enough to show \eqref{stepreqnvdvdeltamplusone}.
\end{minipage}

$\qed$ Theorem \ref{sequencetoQ}.

\begin{minipage}{\textwidth}
\begin{theorem}
\label{ordersiblings}
The rational key corresponding to a node is less than the rational key corresponding to its next sibling.\\[2mm]

FOR ALL
\begin{align*}
& m \in \mathbb{N}\\
\intertext{AND}
& nv_m, dv_m, snv_m, sdv_m, N_1, ..., N_m \in \mathbb{N}
\end{align*}

PROVIDED
\begin{equation}
\label{ordersiblingshyp}
\Bigl [
\begin{smallmatrix} nv_m & snv_m\\ dv_m & sdv_m \end{smallmatrix}
\Bigr ]
=
\Bigl [
\begin{smallmatrix} 0 & 1\\ 1 & 0 \end{smallmatrix}
\Bigr ]
\prod_{k=1}^{m} 
\Bigl [
\begin{smallmatrix} 1 & 1\\ N_{k} & (N_{k} + 1) \end{smallmatrix}
\Bigr ]
\end{equation}

HOLDS
\begin{equation}
\label{ordersiblingsreq}
\frac{nv_m}{dv_m} < \frac{snv_m}{sdv_m}
\end{equation}

\end{theorem}
\end{minipage}

\subsection*{Proof of Theorem \ref{ordersiblings}}

\begin{minipage}{\textwidth}
Recalling calculation \eqref{detequalsunity}, given
\eqref{ordersiblingshyp}, it follows that:
\begin{equation}
\label{detequalsunityagain}
nv_m \times sdv_m - dv_m \times snv_m = -1
\end{equation}
\end{minipage}

\begin{minipage}{\textwidth}
Since each of $dv_m$ and $sdv_m$ is strictly positive, it follows that:
\begin{equation}
\label{detequalsunityagaindivided}
\frac{nv_m}{dv_m} - \frac{snv_m}{sdv_m} = \frac{-1}{dv_m \times sdv_m}
\end{equation}
\end{minipage}

\begin{minipage}{\textwidth}
Or:
\begin{equation}
\label{detequalsunityagaindividedrearranged}
\frac{nv_m}{dv_m} + \frac{1}{dv_m \times sdv_m} = \frac{snv_m}{sdv_m}
\end{equation}
\end{minipage}

From which \eqref{ordersiblingsreq} is immediate.

$\qed$ Theorem \ref{ordersiblings}.

\begin{minipage}{\textwidth}
\begin{theorem}
\label{collectchildren}
The rational key corresponding to a child node lies strictly between the rational key of its parent and the rational key of its parents next sibling.\\[2mm]

FOR ALL
\begin{align*}
& m \in \mathbb{N}\\
\intertext{AND}
& nv_m, dv_m, snv_m, sdv_m, nv_{m+1}, dv_{m+1}, snv_{m+1}, sdv_{m+1}, N_1, ..., N_{m+1} \in \mathbb{N}
\end{align*}

PROVIDED
\begin{equation}
\label{collectchildrenhyp}
\Bigl [
\begin{smallmatrix} nv_m & snv_m\\ dv_m & sdv_m \end{smallmatrix}
\Bigr ]
=
\Bigl [
\begin{smallmatrix} 0 & 1\\ 1 & 0 \end{smallmatrix}
\Bigr ]
\prod_{k=1}^{m} 
\Bigl [
\begin{smallmatrix} 1 & 1\\ N_{k} & (N_{k} + 1) \end{smallmatrix}
\Bigr ]
\end{equation}

\begin{equation}
\label{collectchildrenstephypplusone}
\Bigl [
\begin{smallmatrix} nv_{m+1} & snv_{m+1}\\ dv_{m+1} & sdv_{m+1} \end{smallmatrix}
\Bigr ]
=
\Bigl [
\begin{smallmatrix} 0 & 1\\ 1 & 0 \end{smallmatrix}
\Bigr ]
\prod_{k=1}^{m+1} 
\Bigl [
\begin{smallmatrix} 1 & 1\\ N_{k} & (N_{k} + 1) \end{smallmatrix}
\Bigr ]
\end{equation}

HOLDS
\begin{equation}
\label{collectchildrenreq}
\frac{nv_m}{dv_m} < \frac{nv_{m+1}}{dv_{m+1}} < \frac{snv_m}{sdv_m}
\end{equation}

\end{theorem}
\end{minipage}

\subsection*{Proof of Theorem \ref{collectchildren}}

To prove \eqref{collectchildrenreq} it is required to prove
\begin{equation}
\label{collectchildrenreqone}
\frac{nv_m}{dv_m} < \frac{nv_{m+1}}{dv_{m+1}}
\end{equation}
and to prove
\begin{equation}
\label{collectchildrenreqtwo}
\frac{nv_{m+1}}{dv_{m+1}} < \frac{snv_m}{sdv_m}
\end{equation}

\begin{minipage}{\textwidth}
Again it is useful to first calculate, using \eqref{collectchildrenstephypplusone} and \eqref{collectchildrenhyp}:
\begin{equation}
\label{collectchildrenstepmplusoneexpand}
\begin{split}
\Bigl [
\begin{smallmatrix} nv_{m+1} & snv_{m+1}\\ dv_{m+1} & sdv_{m+1} \end{smallmatrix}
\Bigr ]
& = \Bigl [
\begin{smallmatrix} nv_m & snv_m\\ dv_m & sdv_m \end{smallmatrix}
\Bigr ] \Bigl [
\begin{smallmatrix} 1 & 1\\ N_{m+1} & (N_{m+1} + 1) \end{smallmatrix}
\Bigr ] \\
&= \Bigl [
\begin{smallmatrix} (nv_m + N_{m+1} \times snv_m) & (nv_m + (N_{m+1} + 1) \times snv_m)\\  (dv_m + N_{m+1} \times sdv_m) & (dv_m + (N_{m+1} + 1) \times sdv_m) \end{smallmatrix}
\Bigr ]
\end{split}
\end{equation}
\end{minipage}

\begin{minipage}{\textwidth}
The requirement \eqref{collectchildrenreqone}:
\begin{equation*}
\begin{split}
& && \frac{nv_m}{dv_m} < \frac{nv_{m+1}}{dv_{m+1}} & \\
& \iff && \frac{nv_m}{dv_m} < \frac{nv_m + N_{m+1} \times snv_m}{dv_m + N_{m+1} \times sdv_m} & \text{using \eqref{collectchildrenstepmplusoneexpand}}\\
& \iff && nv_m \times(dv_m + N_{m+1} \times sdv_m) < (nv_m + N_{m+1} \times snv_m) \times dv_m & \text{all terms $\in \mathbb{N}$}\\
& \iff && N_{m+1} \times nv_m \times sdv_m < N_{m+1} \times snv_m \times dv_m & \text{simplifying}\\
& \iff && nv_m \times sdv_m < snv_m \times dv_m & \text{since $N_{m+1} \in \mathbb{N}$}
\end{split}
\end{equation*}
which follows as a result of Theorem \ref{ordersiblings}. 
\end{minipage}

\begin{minipage}{\textwidth}
The requirement \eqref{collectchildrenreqtwo}:
\begin{equation*}
\begin{split}
& && \frac{nv_{m+1}}{dv_{m+1}} < \frac{snv_m}{sdv_m} & \\
& \iff && \frac{nv_m + N_{m+1} \times snv_m}{dv_m + N_{m+1} \times sdv_m} < \frac{snv_m}{sdv_m} & \text{using \eqref{collectchildrenstepmplusoneexpand}}\\
& \iff && (nv_m + N_{m+1} \times snv_m) \times sdv_m < snv_m \times (dv_m + N_{m+1} \times sdv_m) & \text{all terms $\in \mathbb{N}$}\\
& \iff && nv_m \times sdv_m + N_{m+1} \times snv_m \times sdv_m < snv_m \times dv_m + N_{m+1} \times snv_m \times sdv_m & \text{simplifying}\\
& \iff && nv_m \times sdv_m < snv_m \times dv_m &
\end{split}
\end{equation*}
which follows as a result of Theorem \ref{ordersiblings}. 
\end{minipage}

$\qed$ Theorem \ref{collectchildren}.

\section{Acknowledgements}

Thanks to Evan Jones for his review and suggestions.

\bibliographystyle{alpha}
\bibliography{UsingRationalNumbersToKeyNestedSets}

\begin{thebibliography}{Tro05}

\bibitem[Cel04]{983361}
Joe Celko.
\newblock {\em Joe Celko's SQL for Smarties: Trees and Hierarchies}.
\newblock Morgan Kaufmann Publishers Inc., San Francisco, CA, USA, 2004.

\bibitem[Tro04]{DBLP:journals/corr/cs-DB-0402051}
Vadim Tropashko.
\newblock Nested {I}ntervals {T}ree {E}ncoding with {C}ontinued {F}ractions.
\newblock {\em CoRR}, cs.DB/0402051, 2004.

\bibitem[Tro05]{DBLP:journals/sigmod/Tropashko05}
Vadim Tropashko.
\newblock {N}ested intervals tree encoding in {SQL}.
\newblock {\em SIGMOD Record}, 34(2):47--52, 2005.

\end{thebibliography}

\end{document}